\documentclass[manuscript,screen]{acmart}

\AtBeginDocument{%
  \providecommand\BibTeX{{%
    Bib\TeX}}}

\setcopyright{acmlicensed}
\copyrightyear{2018}
\acmYear{2018}
\acmDOI{XXXXXXX.XXXXXXX}

\acmConference[Conference acronym 'XX]{Make sure to enter the correct
  conference title from your rights confirmation emai}{June 03--05,
  2018}{Woodstock, NY}
\acmISBN{978-1-4503-XXXX-X/18/06}

\usepackage{hyperref}

\begin{document}

\title{Enhancing POI Recommendation through Global Graph Disentanglement with POI Weighted Module}

\author{Pei-Xuan Li}
\authornote{Both authors contributed equally to this research.}
\email{n28121107@gs.ncku.edu.tw}
\orcid{0000-0002-6014-4191}
\affiliation{%
  \institution{Department of Electrical Engineering, National Cheng Kung University}
  \city{Tainan}
  \country{Taiwan(R.O.C)} 
}

\author{Wei-Yun Liang}
\authornotemark[1]
\email{n26110647@gs.ncku.edu.tw}
\affiliation{%
  \institution{Department of Electrical Engineering, National Cheng Kung University}
  \city{Tainan}
  \country{Taiwan(R.O.C)} 
}

\author{Fandel Lin}
\email{fandel.lin@usc.edu}
\orcid{0000-0001-7024-2476}
\affiliation{%
  \institution{USC Information Sciences Institute}
  \department{University of Southern California}
  \streetaddress{4676 Admiralty Way, Suite 1001}
  \city{Marina del Rey}
  \state{CA}
  \country{USA}
  \postcode{90292}
}

\author{Hsun-Ping Hsieh}
\authornote{Corresponding author}
 \affiliation{%
\institution{Department of Electrical Engineering, Academy of Innovative Semiconductor and Sustainable Manufacturing, National Cheng Kung University }
  \city{Tainan}
\country{Taiwan(R.O.C)}}
\email{hphsieh@mail.ncku.edu.tw}
\orcid{0000-0001-6924-1337}

\renewcommand{\shortauthors}{Liang and Li et al.}

\begin{abstract}
  Next point of interest (POI) recommendation primarily predicts future activities based on users' past check-in data and current status, providing significant value to users and service providers. We observed that the popular check-in times for different POI categories vary. For example, coffee shops are crowded in the afternoon because people like to have coffee to refresh after meals, while bars are busy late at night. However, existing methods rarely explore the relationship between POI categories and time, which may result in the model being unable to fully learn users' tendencies to visit certain POI categories at different times. Additionally, existing methods for modeling time information often convert it into time embeddings or calculate the time interval and incorporate it into the model, making it difficult to capture the continuity of time. Finally, during POI prediction, various weighting information is often ignored, such as the popularity of each POI, the transition relationships between POIs, and the distances between POIs, leading to suboptimal performance. To address these issues, this paper proposes a novel next POI recommendation framework called \textbf{G}raph \textbf{D}isentangler with \textbf{P}OI \textbf{W}eighted Module (GDPW). This framework aims to jointly consider POI category information and multiple POI weighting factors. Specifically, the proposed GDPW learns category and time representations through the Global Category Graph and the Global Category-Time Graph. Then, we disentangle category and time information through contrastive learning. After prediction, the final POI recommendation for users is obtained by weighting the prediction results based on the transition weights and distance relationships between POIs. We conducted experiments on two real-world datasets, and the results demonstrate that the proposed GDPW outperforms other existing models, improving performance by 3\% \text{to} 11\%\text{.}

\end{abstract}



\received{20 February 2007}
\received[revised]{12 March 2009}
\received[accepted]{5 June 2009}

\maketitle

\section{Introduction}

Location-Based Services (LBS) have advanced significantly due to the proliferation of GPS systems and ride-sharing applications. APPs like Gowalla, Yelp, and Foursquare enable users to share experiences at points of interest (POIs), generating large amounts of spatio-temporal data. These kinds of data can be leveraged to create the next POI recommendation system \cite{5}, which predicts a user’s destination based on their check-in history. Such a system can assist users in exploring new places and enhancing their overall experience. 

Traditional methods \cite{9, 19} capture the transition relationships between POIs. Later on, recurrent neural networks (RNNs) and attention-based models emerged. The former approaches, including Long Short-Term Memory (LSTM) and Gated Recurrent Unit (GRU), have memory mechanisms that retain both long-term and short-term information while integrating temporal and spatial information. The latter approaches use Transformer \cite{23} extensively to learn the connection between check-ins \cite{30,33}. Along with utilizing sequential models to capture information, graph-based methods can take into account the geographic relationships between POIs and use graph neural networks (GNNs) to learn POI representations. Given the superior performance of graph-based methods in existing research, this paper centers on employing a graph-based approach. Although previous graph-based methods have achieved promising results, they still suffer from limitations that may lead to suboptimal performance. One notable limitation is their tendency to overlook the relationship between POI categories and time. 

\begin{figure}[h]
  \centering
  \includegraphics[width=\linewidth]{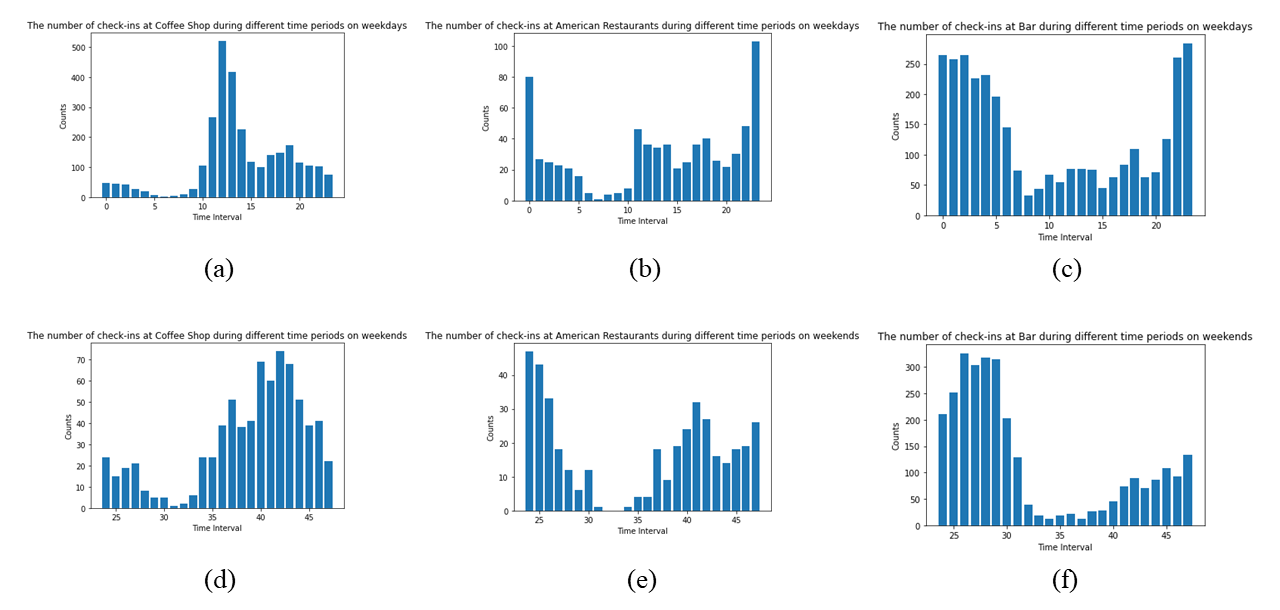}
  \caption{Analysis of check-in numbers for POI categories at different time slots.}
  \Description{Analysis of check-in numbers for POI categories at different time slots.}
  \label{fig:category time analysis}
\end{figure}

Firstly, POI categories and time are strongly correlated. For instance, Figure \ref{fig:category time analysis} analyzes the number of check-ins for three different categories across various time periods, with separate calculations for weekdays and weekends. Figure \ref{fig:category time analysis} (a) and Figure \ref{fig:category time analysis} (d) show the "Coffee Shop" category, where check-ins are concentrated around midday and afternoon, indicating that people might habitually drink coffee to stay alert after lunch or meet friends at coffee shops. Figure \ref{fig:category time analysis} (b) and Figure \ref{fig:category time analysis} (e) show the "American Restaurants" category, with check-ins peaking during dinner hours. Figure \ref{fig:category time analysis} (c) and Figure \ref{fig:category time analysis} (f) show the "Bar" category, with check-ins concentrated in the late-night hours. The data analysis results indicate that behavioral patterns can vary significantly across different times of the day. However, existing methods \cite{30,33,36,38} rarely study the relationship between POI categories and time, which may result in models failing to learn the temporal dependencies of user behaviors. Secondly, temporal information is primarily incorporated by converting POI check-in times into 24 or 48 time slots, or by calculating the time intervals between check-ins and integrating them into the model \cite{31,33}. Although these methods capture temporal information, they only consider the current time context and overlook temporal continuity. Previous research \cite{17} constructed graphs using time information and connected adjacent time nodes, using graph convolutional networks (GCN) \cite{8} to propagate information between categories and time. This method of connecting each time node to adjacent time points may cause the model to process unnecessary temporal information, leading to suboptimal performance. Finally, POIs contain a wealth of information, such as the popularity of each POI, the transition weights (i.e., sequential information) between POIs, and the distances between them. Previous methods \cite{30,33} have not fully leveraged these information, often incorporating only the count of paired check-ins into the model, without considering the inherent popularity of each POI. Besides, \cite{30} use the inverse of the distance between POIs as the edge weight in a geographical graph, but they overlook the significant impact of the last check-in location on predicting the next potential POI.

In this study, we propose the \textbf{G}raph \textbf{D}isentangler with \textbf{P}OI \textbf{W}eighted Module (GDPW) to tackle these challenges. The model consists of three main components: (1) the Category-Time Disentangle Layer,  (2) the POI Weighted Layer, and (3) the Prediction Layer. 
Firstly, to effectively obtain accurate category and time representations and to disentangle the complex category and time information in the original data, we propose the Category-Disentangle Layer. We learn category and time representations by constructing a Global Category Graph and a Global Category-Time Graph. Next, we disentangle category and time information to help the model learn the user's time dependencies.
To consider the rich information of POIs, such as the popularity of POIs, transition weights, and the distance between POIs, we add a module, POI Weighted Layer, to the model. The module incorporates the popularity of POIs when constructing the Global Universal Gravity (UG) Graph, while the Transition Weighted Map (TM) calculates the transition weights between POIs to summarise the sequence information of the user's movement between POIs. In addition, Distance Map (DM) calculates the distance between them.
Finally, the Prediction Layer makes predictions and applies weighting using the DM and TM to the predicted results to summarise all the information.

Our key contributions are summarized as follows:

\begin{itemize}

\item We propose \textbf{G}raph \textbf{D}isentangler with \textbf{P}OI \textbf{W}eighted Module (GDPW), a novel model that integrates POI categories and temporal information into a single graph, enabling category embeddings to capture temporal dependencies more effectively.

\item GDPW employs disentanglement techniques to separate and enhance the learning of both POI category and temporal information, improving the model’s ability to capture distinct influences on user mobility patterns.

\item We introduce a POI weighting mechanism that considers POI popularity, transition weights, and distances, allowing the model to prioritize locations with higher check-ins and those closer to the user, thereby refining recommendation accuracy.

\item Extensive experiments on two real-world datasets demonstrate that GDPW outperforms existing methods, achieving an overall improvement of 3\% to 11\% in both evaluation metrics.

\end{itemize}

The rest of this article is structured as follows: Section \ref{sec:RW} provides a brief overview of related work. In Section \ref{sec:Preliminaries}, we define the next POI recommendation problem and present the notations used. Section \ref{sec:Framework} introduces the proposed GDPW model, detailing the design and functionality of each module. Section \ref{sec:Experiment} presents extensive experiments to assess the effectiveness of GDPW, followed by conclusions and future directions in Section \ref{sec:conclusion}.

\section{Related Works}
\label{sec:RW}
\subsection{Next POI Recommendation}
Next POI recommendation task involves predicting the most likely location a user will visit next by analyzing their recent or past movement patterns. Early research primarily utilized non-deep learning methods such as Markov chains (MC) \cite{2,21,35}, Collaborative Filtering (CF) \cite{6,34}, and Matrix Factorization (MF) \cite{9,10,13,16}. However, these methods have significant deficiencies. They rely on local information, fail to capture overall trends and high-order patterns, and often suffer from sparsity and cold-start problems. 

Nowadays, researchers utilize deep learning models, especially RNNs and LSTMs, to address the next POI recommendation problem, capturing higher-level sequential correlations \cite{12,22,27,28,39}. For instance, ST-RNN \cite{12} utilizes time and distance-specific transition matrices to more effectively capture spatio-temporal contexts. DeepMove \cite{4} utilizes RNNs to learn long-term relationships while employing a historical attention module to identify which historical activities are more important. STAN \cite{14} uses self-attention layers and a spatio-temporal relation matrix to learn POI correlations, demonstrating the importance of non-adjacent POIs for recommendation results. However, these approaches rely solely on sequential models for the next POI recommendation, overlooking the need to learn movement patterns and POI distributions from a global perspective. 

In terms of research on POI categories, previous studies \cite{38} directly transformed POI categories into embeddings integrated into the model without incorporating time-related information. Some methods \cite{30,33,36} convert POI categories and time into separate embeddings, but this approach fails to fully capture the temporal dependencies of user behavior. Additionally, Some methods use personalized approaches \cite{17} to handle the correlation between POI categories and time; however, they consume significant computational resources.

\subsection{Graph-based POI Recommendation}
In recent years, graph-based models have achieved promising results, including GTAG \cite{37}, GE \cite{29}, GSTN \cite{25}, STP-UDGAT \cite{11}, GETNext \cite{33}, and MobGT \cite{30}. Yuan et al. \cite{37} proposed the geographical-temporal influence aware graph (GTAG), composed of user, session, and POI nodes. However, a large number of nodes can make the graph too large, resulting in low processing efficiency. Unlike GTAG, Xie et al. \cite{29} constructed four bipartite graphs, including POI-Region, POI-POI, POI-Time, and POI-Word, to learn geographical, sequential, temporal, and semantic information, respectively. GSTN \cite{25} uses transition-based and distance-based graphs to capture spatial information and designs a Time-LSTM to learn temporal information. STP-UDGAT \cite{11} constructs spatial, temporal, and preference graphs, but using the reciprocal of time as the edge weight may not effectively capture temporal information. GETNext \cite{33} uses check-in data from all users to construct POI transition graphs and employs a Transformer to learn the relationships between sequences. MobGT \cite{30} integrates distinct spatial and temporal graph encoders to capture unique features and global user-location relationships. It also includes a graph-based transformer mobility encoder to extract high-order relationships between POIs. While the above graph-based methods have achieved promising results, none of them simultaneously consider both the popularity and distance of POIs within the graph, and few take temporal continuity into account.

\subsection{Disentangle Representation Learning}
Disentangle Representation Learning \cite{1} seeks to extract distinct explanatory representations from intertwined latent factors. 
For example, MacridVAE \cite{15} employs Variational Auto-Encoders (VAEs) to encode items via disentangled prototypes, aiming to decompose user interactions from both macro and micro levels to capture user preferences regarding different intentions. 
CLSR \cite{40} uses disentanglement techniques to separate long-term and short-term interests. 
DRAN \cite{24} constructs graphs based on distance and transition, and uses the proposed DGCN to model different aspects of POIs while learning POI representations. 
DisenPOI \cite{18} posits that the Geographical Graph and Sequential Graph contain intertwined information, and aims to use disentanglement techniques to aid in learning. CLSPRec \cite{3} captures both long-term and short-term behavior sequences of users, ensuring simplicity while capturing consistent features across different time periods. 
DCHL \cite{lai2024disentangled} disentangles user preferences across collaborative, transitional, and geographical aspects using hypergraph structures, while leveraging cross-view contrastive learning to model cooperative associations.
SD-CEM \cite{jia2024learning} improves POI category embedding by leveraging hierarchical category structures and disentangled mobility sequences, using an attention-based hierarchy-enhanced embedding method and pretext tasks.
CDLRec \cite{yang2024contrastive} leverages contrastive learning and structure causal models to disentangle causal and confounding representations, using uniform data as a supervision signal to enhance unbiased user and item representations, improving debiased recommendation performance.
DIG \cite{qin2022disentangling} proposed a disentangled representation learning approach to separate user interest and geographical factors, leveraging a geo-constrained negative sampling strategy and a geo-enhanced soft-weighted loss function to improve performance.
Different from prior research, we constructed category graphs and category-time graphs, leveraging disentanglement to learn representations of category and time.

\section{Preliminaries}
\label{sec:Preliminaries}
\subsection{Problem Formulation}
Let \( U = \{\mathit{u}_1, \mathit{u}_2, \dots, \mathit{u}_{|U|}\} \) be a set of \(|U|\) users, \( P = \{\mathit{p}_1, \mathit{p}_2, \dots, \mathit{p}_{|P|}\} \) be a set of \(|P|\) POIs and \( C = \{\mathit{c}_1, \mathit{c}_2, \dots, \mathit{c}_{|C|}\} \) be a set of \(|C|\) categories. Each location \( p \in P \) is represented as a tuple \( p = \langle \mathit{lat}, \mathit{lon}, \mathit{cat} \rangle \) of latitude, longitude, and category. Furthermore, let \( T = \{t_1, t_2, \dots, t_{|T|}\} \) represent the check-in times.
\begin{itemize}
\item A check-in can be represented as a tuple \( s = \langle u, p, t \rangle \in S^U \), indicating that user \( u \) visited POI \( p \) at timestamp \( t \).
\item Given a user \( u \in U \), all check-in activities can form a check-in sequence \( S^u = (s_1^u, s_2^u, s_3^u, \dots) \), where \( S^u \) is ordered by timestamp \( t \), and \( s_i^u \) is the \( i \)-th check-in record.
\item Denote the check-in sequence of all users as \( S^U = \{s^{u_1}, s^{u_2}, \dots, s^{u_{|U|}} \} \).
In the realm of data processing, we isolate consecutive check-in records that surpass a specified time interval (e.g., 24 hours) in the check-in sequence \( S^u \) of any user \( u \) into a set of consecutive trajectories, where \( S^u = \{Q_1^u, Q_2^u, \dots \} \).
\item The goal of the next POI recommendation is to predict the next POI that a user \( u \) might visit by learning from their current trajectory and the historical check-in records of all users. It provides a list of recommended POIs.

To be more specific with math, given all users' historical trajectories \( \{Q_i^u\} \) where \( i \in \mathbb{N} \) and \( u \in U \), and a current trajectory \( Q' = (s_1, s_2, \dots, s_k) \) of a specific user \( u \in U \), output the top-\(n\) POIs \( \{s_{(k+1)}^1, s_{(k+1)}^2, \dots, s_{(k+1)}^n\} \) that the user \( u \) is most likely to visit next.

\end{itemize}
\subsection{Construction of Three Global Graphs}
To better learn POI representations, we construct three graphs: Global Category Graph, Global Category-Time Graph, and Global UG Graph, to learn information about category, time, and spatial aspects. All graphs are constructed with a global view.

\subsubsection{Global Category Graph}
We use global category graph to describe the relationships of POI categories.
The directed category-category graph \( G_c = \{V_c, E_c, A_c\} \) is built upon the category of POIs, denoted as \( C \). A directed edge \( e_c = (c_i, c_j) \in E_c \) indicates that there is a check-in from category \( c_i \) to category \( c_j \). We connect each directed category pair based on all users’ trajectories \( Q^U \). The edge weight matrix \( A_c(i,j) \) between each pair represents the frequency of category pair occurrences.

\subsubsection{Global Category-Time Graph}
To capture the relationship between category and time patterns, we propose a Global Category-Time Graph, which is constructed as a heterogeneous graph to record the visiting time for different categories across users. First, we need to convert continuous time to discrete intervals. We discretize check-in times into 48 intervals to differentiate activities in weekdays and weekends: 0 to 23 for weekdays, and 24 to 47 for weekends, respectively. The set of time slots is denoted by \( T_{48} \). The Global Category-Time Graph is denoted as \( G_{ct} \).

In the Global Category-Time Graph, categories and times are represented by category nodes \( c \in C \) and time nodes \( t \in T_{48} \). Here, \( C \) is the category node set, and \( T_{48} \) is the time node set. These two types of nodes are connected by six relations of weighted directed edges: \( E_{cto} \), \( E_{tco} \), \( E_{ctf} \), \( E_{tcf} \), \( E_{ctb} \), and \( E_{tcb} \), where \( E_{(X,Y,R)} \) denotes the sets of edges from nodes in set \( X \) to nodes in set \( Y \), and \( R \) denotes the relationship between \( X \) and \( Y \). Here, \( o \) stands for “original”, \( f \) stands for “forward,” and \( b \) stands for “backward” in relation \( R \).

Next, we will explain how the edge weights are set in the Global Category-Time Graph and the rationale behind this design. To enable the model to learn the concept of continuous time, we divided the edges in the Global Category-Time Graph into three types: original (representing the current time point), forward (representing the previous time point), and backward (representing the next time point).

\begin{itemize}
\item \textbf{Weights of edges in the original relation}
The relationship that “original” aims to capture is the category of the check-in and their corresponding time slots. An edge \( e_{cto} = (c_i, t_i) \in E_{cto} \) indicates that there is a check-in of category \( c_i \) at time \( t_i \). The edge weight between category and time is the frequency of category-time pair occurrences. The adjacency matrix for this relationship is denoted as \( A_{ct}^o \).
\item \textbf{Weights of edges in the forward relation}
The “forward” relationship aims to record the check-in category and the preceding time slot of the check-in. An edge \( e_{ctf} = (c_i, t_i) \in E_{ctf} \) indicates that the check-in category \( c_i \) occurred at the forward time \( t_i \). The edge weight between category and time is the frequency of category-time pair occurrences. The adjacency matrix for this relationship is denoted as \( A_{ct}^f \).
\item \textbf{Weights of edges in the backward relation}
The “backward” relationship aims to record the check-in category and the subsequent time slot of the check-in. An edge \( e_{ctb} = (c_i, t_i) \in E_{ctb} \) indicates that the check-in category \( c_i \) occurred at the backward time \( t_i \). The edge weight between category and time is the frequency of category-time pair occurrences. The adjacency matrix for this relationship is denoted as \( A_{ct}^b \).
\end{itemize}

\begin{figure}[h]
  \centering
  \includegraphics[width=\linewidth]{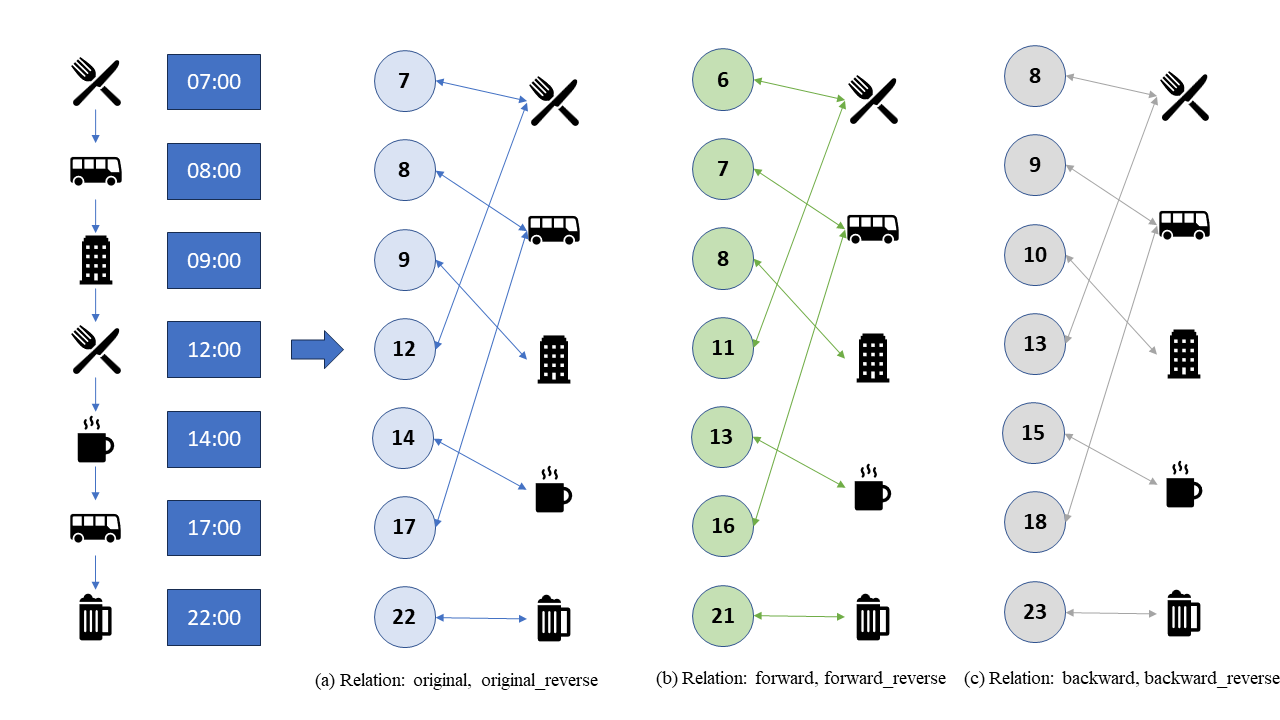}
  \caption{Global Category-Time Graph.}
  \Description{Global Category-Time Graph.}
  \label{fig:global category-time graph}
\end{figure}

The remaining three types of edges are reverse edges, all pointing from time to category. By connecting the current time slot, the previous time slot, and the next time slot, the category can learn the continuity of time. Figure \ref{fig:global category-time graph} illustrates how the Global Category-Time Graph is constructed. For example, when a user visits a restaurant at 7 AM, the original and original\_reverse relations will create an edge between the category 'Restaurant' and the time slot '7' in the Global Category-Time Graph. The forward and forward\_reverse relations will connect the category 'Restaurant' with the previous time slot '6,' while the backward and backward\_reverse relations will link the category 'Restaurant' to the next time slot '8.'

\subsubsection{Global Universal Gravity Graph}

Based on our findings, users tend to prefer POIs that are closer rather than those that are farther away. Additionally, most POIs have fewer than 50 check-ins, indicating that check-ins are concentrated at specific POIs. For example, in the Foursquare New York City dataset, check-ins are focused on landmarks such as Madison Square Park, John F. Kennedy International Airport, Union Square Park, and Times Square. In contrast, in the Foursquare Tokyo dataset, check-ins are concentrated at major train stations like Ikebukuro Train Station, Tokyo Train Station, and Shinjuku Train Station.

To simultaneously consider these two pieces of information, we propose the Global Universal Gravity (UG) Graph \( G_{ug} = \{V_{ug}, E_{ug}, A_{ug}\} \), where the set of nodes \( V_{ug} \) (POIs) is represented by \( P \). An undirected edge between \( p_i \) and \( p_j \) indicates a record transmitted from \( p_i \) to \( p_j \). The edge weight matrix \( A_{ug}(i,j) \) represents the value of Newton’s law of universal gravitation calculated between \( p_i \) and \( p_j \). The formula is as follows:

\begin{equation}
  A_{ug}(i,j) = \frac{GMm}{r^2 + \epsilon}
\end{equation}

where:
\begin{itemize}
    \item \( G \): Frequency of a POI pair \((p_i, p_j)\) across all trajectories.
    \item \( M \): \( p_i \)'s check-in counts in all trajectories.
    \item \( m \): \( p_j \)'s check-in counts in all trajectories.
    \item \( r^2 \): Haversine distance \cite{26} between \( p_i \) and \( p_j \).
    \item \( \epsilon \): A constant added to avoid a zero denominator, which we set to 1.
\end{itemize}

\section{Framework}
\label{sec:Framework}
\subsection{Overview}
\begin{figure}[h]
  \centering
  \includegraphics[width=\linewidth]{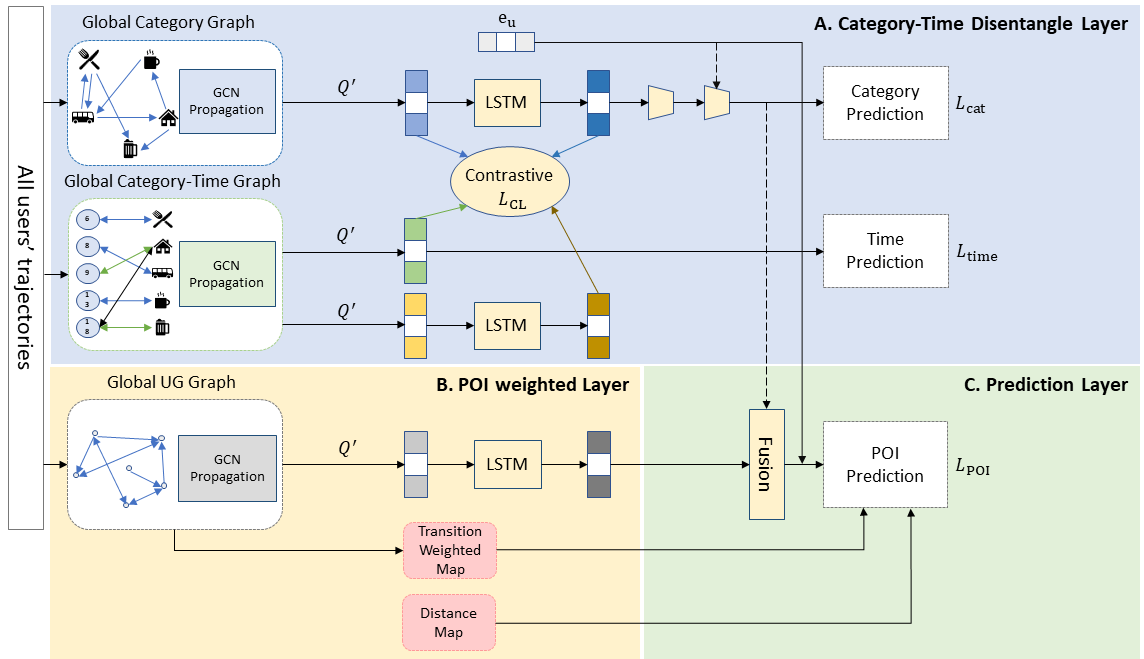}
  \caption{The architecture of GDPW framework.}
  \Description{The architecture of GDPW framework.}
  \label{fig:GDPW}
\end{figure}

In Figure \ref{fig:GDPW}, we present the overall architecture of Graph Disentangler with POI Weighted Module (GDPW), which can be divided into three main parts : (1) \textbf{Category-Time Disentangle Layer} primarily focuses on disentangling category and time representations; (2) \textbf{The POI Weighted Layer} learns representations about POIs and constructs the Transition Weighted Map (TM) and Distance Map (DM) to calculate transition weights and the distances between POIs, converting the rich information in POIs into embeddings; (3) \textbf{The Prediction Layer} conducts predictions and weights the results using the TM and DM to integrate all the information.

\subsection{Category-Time Disentangle Layer}
In this section, we introduce how data propagates on the Global Category Graph and Global Category-Time Graph by graph convolutional networks, and disentangle category representations and time representations through contrastive learning.
\subsubsection{Global Category Graph Propagations}

In this part, we use graph convolutional networks (GCN) \cite{8} to obtain the POI category representations.
Given the Global Category Graph \( G_c \), where the nodes consist of POI categories and the edge weights are defined by the frequency of category transitions across all users, the subsequent step is to learn the representation of categories. To accomplish this, we employ a GCN to learn POI representations, which is also the predominant method used for learning in most graph-based POI recommendation models. We first compute the Random Walk normalized Laplacian as:

\begin{equation}
    \tilde{L} = I - D_c^{-1} A_c,
\end{equation}
where \( A_c \) is the adjacency matrix, and \( D_c \) is the degree matrix of \( G_c \). Next, let \( H^{(0)} = X_c \) be the input node feature matrix of the GCN, where \( X_c \) denotes features about the total number of data entries for this category and the total number of stores for this category. The following formula is the GCN propagation method:

\begin{equation}
H^{(l+1)} = \sigma\left(\tilde{L} H^{(l)} W_c^{(l)} + b_c^{(l)}\right),\
\label{eq4_2}
\end{equation}
where \( H^{(l)} \) is the embedding of the \( l \)-th layer, \( W_c^{(l)} \) is the learnable weight matrix of layer \( l \), \( b_c^{(l)} \) is the bias term of layer \( l \), and \( \sigma \) is the activation function ELU. After passing through an \( L \)-layer GCN, we then obtain output category representations as \( e_c \). Therefore, for each category in each trajectory \( Q' \), it is represented by a list of representations \( E_c^c = \left( e_1^{'c}, \dots, e_k^{'c} \right) \).

\subsubsection{Global Category-Time Graph Propagations}
The Global Category-Time Graph is a heterogeneous graph with six relations of edges, including original, forward, backward, \text{original\_reverse}, \text{forward\_reverse}
 and \text{backward\_reverse}. For each relation, we utilize GCN for learning and processing each type of relationship using the Random Walk Laplacian. We allow messages from source nodes to propagate along different relationships to destination nodes and aggregating information from different relationships to update features for the same target node. The formula is as follows:
 
 \begin{equation}
     h_{dst}^{(l+1)} = \text{AGG}_{r \in R} \left[ r_{dst} = \text{dst} \left( f_r \left( g_r, h_{r_{src}}^{(l)}, h_{r_{dst}}^{(l)} \right) \right) \right],
 \end{equation}
 where \( \text{dst} \) is the destination node, \( f_r \) is the GCN function applied to each relationship as detailed in Eq.~\ref{eq4_2}, and \( \text{AGG} \) is the aggregation function applied at the destination node. After passing through the GCN, we obtain the category representations \( e_{ct_c} \) and time representations \( e_{ct_t} \). For each trajectory \( Q' \), we obtain category representations \( E_{ct}^c = \left( e_1^{'(ct_c)}, \dots, e_k^{'(ct_c)} \right) \) and time representations \( E_{ct}^t = \left( e_1^{'(ct_t)}, \dots, e_k^{'(ct_t)} \right) \).

\begin{figure}[h]
  \centering
  \includegraphics[width=\linewidth]{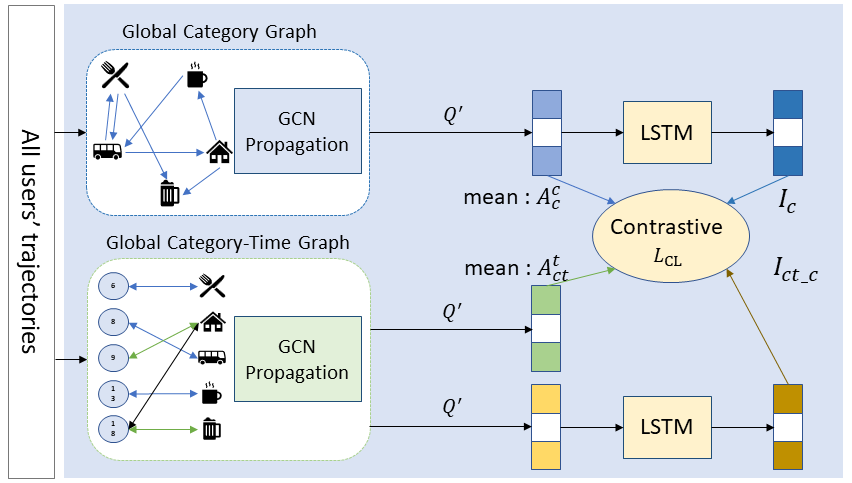}
  \caption{Category-Time Disentangler.}
  \Description{Category-Time Disentangler.}
  \label{fig4_2}
\end{figure}

\subsubsection{Category-Time Disentangler}

Both category representations \( e_c \) from the Global Category Graph and \( e_{ct_c} \) from the Global Category-Time Graph are associated with category transitions and time. To enable the model to learn the relationships between different POI categories and time, we use a disentangling method to aid in the learning process.

Inspired by previous work on self-supervised disentanglement learning \cite{17,18,40}, we use contrastive learning methods to perform disentanglement operations.
Contrastive learning is a self-supervised learning technique where a model learns to differentiate between similar and dissimilar data points. The core idea is to bring similar samples (positive pairs) closer in the learned representation space while pushing dissimilar samples (negative pairs) apart.

The Category-Time Disentangler is illustrated in Figure ~\ref{fig4_2}. 
We generate proxies from the graph, with each proxy carrying information about its corresponding category or time features from the graph. Two proxies are generated by averaging \( E_{ct}^c \) and \( E_{ct}^t \) as follows:

\begin{equation}
    a_c^c = \frac{1}{k} \sum_{i=1}^k e_i^{'c},
\end{equation}
\begin{equation}
    a_{ct}^t = \frac{1}{k} \sum_{i=1}^k e_i^{'(ct_t)},
\end{equation}
where \( a_c^c \) carries category transition information, while \( a_{ct}^t \) carries category information related to time.

Regarding the two sets of representations to be learned, we don't directly disentangle them. Instead, they first go through two different sets of LSTM to learn the temporal relationships and get hidden states \( h_n^c \) and \( h_n^{(ct_c)} \) as follows:

\begin{equation}
    h_n^c = \text{LSTM}\left(e_i^{'c}, h_{(i-1)}^c\right), \quad i \in \{1, 2, \ldots, k\},
\end{equation}
\begin{equation}
    h_n^{(ct_c)} = \text{LSTM}\left(e_i^{'(ct_c)}, h_{(i-1)}^{(ct_c)}\right), \quad i \in \{1, 2, \ldots, k\},
\end{equation}
After obtaining hidden states, we project the hidden states and proxies into another dimension according to previous work \cite{17, 18} to reduce noise, as follows:
\begin{equation}
    I_c = W_1 h_n^c,
\end{equation}
\begin{equation}
    I_{(ct_c)} = W_2 h_n^{(ct_c)},
\end{equation}
\begin{equation}
    A_c^c = W_3 a_c^c,
\end{equation}
\begin{equation}
    A_{ct}^t = W_4 a_{ct}^t,
\end{equation}
where \( W_1 \), \( W_2 \), \( W_3 \), and \( W_4 \) are linear transformations with trainable parameters.

We use contrastive learning to disentangle the information between proxies and given representations. In other words, we treat representations from the same graph as positive samples and representations from different graphs as negative samples \cite{17, 18}. Then, we employ contrastive learning loss with the following formulation:
\begin{equation}
    L_{CL} = F(A_c^c, I_c, I_{(ct_c)}) + F(A_{ct}^t, I_{(ct_c)}, I_c),
    \label{eq_loss_cl}
\end{equation}
where \( \odot \) denotes the inner product and \( F(\cdot) \) denotes the Bayesian Personalized Ranking loss \cite{17, 18, 20, 40}. The formula is as follows:
\begin{equation}
    F(\text{pro}, \text{pos}, \text{neg}) = \text{Softplus}\left(\langle \text{pro} \odot \text{neg} \rangle - \langle \text{pro} \odot \text{pos} \rangle \right).
\end{equation}
As mentioned previously, we expect the the global category graph representation to be similar to its proxy vectors as well as dissimilar to global category-time graph representation, and vice versa.

\subsubsection{Auxiliary Prediction}
To help the model learning more effectively, we use the above information to predict category and time as auxiliary task after learning and disentangling the category and time representations.
To capture each user's behavior, we also consider user identification by projecting the information into a low-dimensional vector to obtain user embeddings \( e_u \).

For the category prediction part, we concatenate hidden states obtained from LSTM \( h_n^c \), \( h_n^{(ct_c)} \) and user embeddings \( e_u \). Then, we adopt multilayer perceptron models (MLP) as the prediction layer to predict the POI category. The formulas are as follows:

\begin{equation}
    e_{\text{cat}} = W_{\text{fuse1}} \left( h_n^c \| h_n^{(ct_c)} \right) + b_{\text{fuse1}},
\end{equation}
\begin{equation}
    \hat{y}_{\text{cat}} = W_{\text{cat}} \left( e_{\text{cat}} \| e_u \right) + b_{\text{cat}},
\end{equation}
where \( W_{\text{fuse1}} \), \( W_{\text{cat}} \) and \( b_{\text{fuse1}} \), \( b_{\text{cat}} \) are learnable weight and bias, respectively.
In the time prediction aspect, we use the time representations from the Global Category-Time Graph \( E_{ct}^t \) to perform the prediction. Additionally, since users' behavior may vary throughout the day, we encode the day of the week into week embeddings \( e_w \) and perform time prediction. The prediction formula is as follows:
\begin{equation}
\hat{y}_{\text{time}} = W_{\text{time}} \left( E_{ct}^t \| e_{\text{week}} \right) + b_{\text{time}},
\end{equation}
where \( W_{\text{time}} \) and \( b_{\text{time}} \) are learnable weight and bias.

\subsection{POI Weighted Layer}
To make the most of the wealth of information in POI data, we learn the POI representations using the Global UG Graph, and then we utilize the Transition Weighted Map and Distance Map to weight the results.

\subsubsection{Global UG Graph Propagations}

Once the category and time information have been disentangled, we move on to capturing the correlations between POIs using GCN \cite{8}. Given the undirected Global UG Graph \( G_{ug} \), with edge weight defined by Newton's law of universal gravitation, we can compute the normalized Laplacian matrix as follows:
\begin{equation}
    \tilde{L} = I - D_{ug}^{-1/2} A_{ug} D_{ug}^{-1/2},
\end{equation}
where \( A_{ug} \) represents the adjacency matrix and \( D_{ug} \) represents the degree matrix of \( G_{ug} \).

Let \( H^{(0)} = X_{ug} \) be the input node feature matrix for the GCN, where \( X_{ug} \) consists of latitude, longitude, category, and frequency. The following is the GCN propagation method:

\begin{equation}
    H^{(l+1)} = \sigma \left( \tilde{L} H^{(l)} W_{ug}^{(l)} + b_{ug}^{(l)} \right),
\end{equation}
where \( H^{(l)} \) is the information of the \( l \)-th layer, \( W_{ug}^{(l)} \) is the weight matrix of layer \( l \), \( b_{ug}^{(l)} \) is the bias term of layer \( l \), and \( \sigma \) is the activation function ELU. After going through \( L \)-layer GCN, we obtain UG representations \( e_{ug} \). For each trajectory \( Q' \), we obtain POI representations \( E_p^p = \left( e_1^p, \ldots, e_k^p \right) \).

The same approach applies to both the Global Category Graph and the Global Category-Time Graph. After propagation, we utilize an LSTM to learn the temporal relationships of \( E_p^p \):
\begin{equation}
    h_n^p = \text{LSTM} \left( e_i^p, h_{(i-1)}^p \right), \quad i \in \{1, 2, \ldots, k\},
\end{equation}
where \( h_n^p \) is the hidden state of \( E_p^p \) obtained from LSTM.

\subsubsection{Transition Weighted Map}



While the Global UG Graph can learn POI representations, we have observed that the previous POI visited significantly influences the selection of the next POI. This sequential dependency plays a crucial role in user movement patterns and should be considered in the model. To incorporate this dependency, we employ the Transition Attention Map from GETNext \cite{33}. In GETNext, the authors argue that using a graph alone can only capture potential user movement information, so they introduced the Transition Attention Map to enhance this information. Similarly, we aim to use this module to calculate the transition weight between POIs. Given the input node features \( X_{ug} \) and \( G_{ug} \), we compute the Transition Weighted Map (TM) as follows:
\begin{equation}
    \varphi_1 = (X_{ug} \times W_{tp1}) \times a_1 \in \mathbb{R}^{|P| \times 1},
\end{equation}
\begin{equation}
    \varphi_2 = (X_{ug} \times W_{tp2}) \times a_2 \in \mathbb{R}^{|P| \times 1},
\end{equation}
\begin{equation}
    TM = (\varphi_1 \times 1^T + 1 \times \varphi_2^T) \odot (\tilde{L} + J_N) \in \mathbb{R}^{|P| \times |P|},
\end{equation}
where \( W_{tp1} \) and \( W_{tp2} \) are two transformation matrices, \( a_1 \) and \( a_2 \) are learnable vectors, \( 1 \) is an all-ones vector, \( J_N \) is the matrix of ones, \( \odot \) denotes element-wise multiplication.

\subsubsection{Distance Map}

Based on our observations, we found that the spatial interval between two POI transitions is often within a certain range. Although we have a Global UG Graph and a Transition Weighted Map to learn POI spatial information, these methods do not effectively narrow the prediction scope. To address this issue, we propose the Distance Map, which is a statistical method that calculates the actual distance using the Haversine formula and applies a Gaussian kernel function to limit the value to the range \([0,1]\).

The Distance Map \( DM(i,j) \) is defined as:
\begin{equation}
DM(i,j) = 
\begin{cases} 
\exp\left(-\frac{dis(i,j)^2}{2\sigma^2}\right), & 0 < dis(i,j) < \Delta d \\
0, & \text{otherwise}
\end{cases}
\end{equation}
where \( dis(i,j) \) is the distance between POIs \( p_i \) and \( p_j \), \( \sigma \) is the bandwidth parameter of the Gaussian kernel function, which is set to 1. This parameter controls the spread or smoothness of the Gaussian kernel function. A smaller value of \( \sigma \) leads to faster decay, meaning that only POIs that are close together will exhibit significant affinity.

\subsection{Prediction Layer}
\subsubsection{POI Prediction}

After obtaining the POI hidden states \( h_n^p \) and category representations \( e_{cat} \), we employ attention mechanisms to dynamically learn the weights of these two representations, allowing us to capture information more effectively. The formulas are as follows:

The attention weights for category representations are computed as:
\begin{equation}
    W_n^c = \text{softmax}(e_{cat})_i = \frac{\exp(e_{cat})_i}{\sum_{j=1}^k \exp(e_{cat})_j},
\end{equation}

The attention weights for POI representations are computed as:
\begin{equation}
    W_n^{poi} = \text{softmax}(h_i^p) = \frac{\exp(h_i^p)}{\sum_{j=1}^k \exp(h_j^p)},
\end{equation}

In the POI prediction part, we fuse POI representations \( e_{poi} \) and user embeddings \( e_u \) to get the POI prediction as follows:
\begin{equation}
    \hat{y}_{poi} = W_{poi} (e_{poi} \| e_u) + b_{poi},
\end{equation}
where \( W_{poi} \) are learnable weights and \( b_{poi} \) is the bias term. The POI representation \( e_{poi} \) is obtained by adding \( r_c \) and \( r_{poi} \) as follows:
\begin{equation}
    e_{poi} = r_c + r_{poi},
\end{equation}
\begin{equation}
    r_c = \sum_{i=1}^k W_i^c e_{cat},
\end{equation}
\begin{equation}
    r_{poi} = \sum_{i=1}^k W_i^{poi} h_i^p.
\end{equation}

Additionally, we combine this POI recommendation result with the Transition Weighted Map (TM) and Distance Map (DM). By applying these maps to the weighting of the results, we account for the transition weight and distance relationships between POIs. Therefore, the final result is:
\begin{equation}
    \hat{Y}_{(poi,i)} = y_{(poi,i)} + TM_i^{(p_k)} + DM_i^{(p_k)},
\end{equation}
where \( TM^{(p_k)} \) is the \( p_k \)-th row of the Transition Weighted Map and \( DM^{(p_k)} \) is the \( p_k \)-th row of the Distance Map. Here, \( p_k \) stands for the POI of check-in \( s_k \).

\subsubsection{Loss functions}

We employ cross-entropy as the loss function for POI prediction \(L_{poi}\) and category prediction \(L_{cat}\). For time prediction, inspired by \cite{33}, we use Mean Squared Error (MSE) as the loss function. The total loss is obtained by summing up these losses with the contrastive learning loss \(L_{CL}\) obtained from Eq.~\ref{eq_loss_cl}. The calculations for POI loss, category loss, time loss, and the total loss are as follows:
\begin{equation}
    L_{poi} = -\sum_{i=1}^P Y_{poi} \log(Y_{poi}),
\end{equation}
\begin{equation}
    L_{cat} = -\sum_{i=1}^C y_{cat} \log(\hat{y}_{cat}),
\end{equation}
\begin{equation}
    L_{time} = \frac{1}{n} \sum_{i=1}^n (\hat{y}_{time} - y_{time})^2,
\end{equation}
\begin{equation}
    L = L_{poi} + L_{cat} + L_{time} + L_{CL},
\end{equation}
where \(\hat{Y}_{poi}\), \(\hat{y}_{cat}\), and \(\hat{y}_{time}\) represent the predicted values of POI, category, and time, respectively, while \(Y_{poi}\), \(y_{cat}\), and \(y_{time}\) represent the ground truth values of POI, category, and time.

\section{Experimental Results}
\label{sec:Experiment}
\subsection{Experiments Setup}
\subsubsection{Dataset}
In our experiments, we utilized two real-world datasets, Foursquare-NYC \cite{32} and Foursquare-TKY \cite{32}, to evaluate the effectiveness of our approach.

These datasets encompass check-ins recorded in New York City (NYC) and Tokyo (TKY) during the period from 12 April 2012 to 16 February 2013. Each entry in these datasets comprises User identification, POI identification, category identification, timestamp, latitude, and longitude.

We preprocessed the datasets according to the settings of previous related work GETNext \cite{33}. We excluded POIs with fewer than 10 check-in records and users with fewer than 10 check-ins from both datasets. Subsequently, within each user's check-in sequence, we segmented the sequence at consecutive check-ins separated by time intervals exceeding 24 hours, thereby creating the user's check-in trajectories. Trajectories with a check-in length of less than 3 were excluded from the dataset. We partitioned the dataset into training, validation, and testing sets. The first 80\% of the trajectories were allocated to the training set, the middle 10\% to the validation set, and the remaining 10\% to the testing set. 

The Global Category Graph and Global Category-Time Graph are generated from the training set, while the Global UG Graph is created using all POIs. The detailed statistics of the two datasets are presented in Table \ref{tab:foursquare}.

\begin{table}
  \caption{The statistics of the Foursquare dataset.}
  \label{tab:foursquare}
  \begin{tabular}{lcc}
    \toprule
    &NYC&TKY\\
    \midrule
    \# User & 1,083 & 2,293\\
    \# POI & 5,130 & 7,870\\
    \# Category & 208 & 190\\
    \# Check-in	& 104,877 & 361,346\\
    \# Trajectory & 15,992 & 50,680\\
  \bottomrule
\end{tabular}
\end{table}

\subsubsection{Baselines}

We employ the following baselines for comparison with GDPW:
\begin{itemize}
    \item MF \cite{9} is a classical method that decomposes a matrix and learns the latent representations of users and POIs.
    \item LSTM \cite{7} is a variant model based on Recurrent Neural Networks (RNNs), capable of simultaneously handling long and short-term dependencies.
    \item STGCN \cite{39} enhances the conventional LSTM by incorporating spatial gates and temporal gates, while also employing coupled input and forget gates for improved efficiency.
    \item PLSPL \cite{28} utilizes an attention mechanism to learn long-term interests and employs LSTM to capture short-term interests. These are then combined using personalized weights.
    \item GSTN \cite{25} is a graph-based recommendation model that learn the spatial-temporal similarities between POIs.
    \item STAN \cite{14} uses an attention mechanism to leverage spatiotemporal information from check-in sequences, capturing the interactions between non-adjacent POIs.
    \item CLSPRec \cite{3} uses the same Transformer model for both long and short-term modeling and utilizes contrastive learning to distinguish different users' travel preferences.
    \item GETNext \cite{33} employs a global trajectory flow map with a Transformer to predict the next POI while simultaneously addressing the cold start problem.
\end{itemize}

\subsubsection{Evaluation Metrics}

Following \cite{33}, we utilize accuracy@k (\(\text{Acc@k}\)) and mean reciprocal rank (\(\text{MRR}\)) to evaluate the model. \(\text{Acc@k}\) measures the proportion of correctly predicted items among the top-\(k\) recommended items. To gain a better understanding of the position of the correct label, we also adopt \(\text{MRR}\), which measures the quality of recommendations by considering the reciprocal of the rank of the first relevant item in the list.

Given a dataset with \(m\) samples, define \(\text{Acc@k}\) and \(\text{MRR}\) as:
\begin{equation}
    \text{Acc@k} = \frac{1}{m} \sum_{i=1}^{m} \left( 1 \text{ if rank} \leq k \text{ else } 0 \right)
\end{equation}
\begin{equation}
    \text{MRR} = \frac{1}{m} \sum_{i=1}^{m} \frac{1}{\text{rank}}
\end{equation}
Both evaluation metrics indicate that a higher score represents better model performance.

\subsubsection{Experiments Settings}

The key hyperparameter settings in our model are listed below: We employ Adam as the optimizer with a learning rate of \(\text{lr} = 0.0002\). The dimensions of all embeddings are set to \(d = 64\), and the number of GCN layers \(l\) is set to 2. We set \(\Delta d\) in the Distance Map to 5 km.

\subsection{Performance Comparison}

Table \ref{tab:results} shows the comparison between GDPW and baseline models on two datasets. The results show that our model shows improvements on both datasets compared to the baselines. In the NYC dataset, Acc@1 increased by roughly 6\% and MRR by around 3\% compared to the best baseline, GETNext. In the TKY dataset, Acc@1 saw an improvement of about 11\%, with MRR rising by approximately 6\% compared to GETNext.

Overall, our model performs better on the NYC dataset than on the TKY dataset. This may be because the distribution of data in the Tokyo dataset makes it difficult to capture user mobility characteristics, as check-ins are concentrated in specific categories (e.g., Train Station). 
However, in both datasets, our GDPW outperforms all the comparison methods, proving that our proposed model is effective in both distributions as well as its robustness.

We can observe that GDPW shows a greater improvement in Acc@1 and MRR compared to Acc@5 and Acc@10. This is because our model applies significant weighting to POIs, which confines them within a certain range. When the target POI falls outside this range, our model finds it more challenging to make accurate predictions. 

Nonetheless, GDPW outperforms the best baseline models on all datasets. It's challenging for MF to learn the complex transitions between check-ins. Models such as STGCN, PLSPL, STAN, and CLSPRec, which are based on LSTM or attention mechanisms, struggle to learn spatial dependencies. On the other hand, GSTN only learns time, space, and transition information, making it difficult to capture high-level contextual information. Among the baseline models, GETNext is the best performer. Compared to GETNext, GDPW accounts for the continuity of time and uses disentanglement to help learn category and time information. Additionally, it applies weighting based on POI popularity, transition weight, and distance relationships.
\begin{table}[h]
\caption{Performance comparison on two datasets.}
\centering
\begin{tabular}{ccccccccc}
    \toprule
    \textbf{Dataset}& \multicolumn{4}{c}{\textbf{NYC}} & \multicolumn{4}{c}{\textbf{TKY}} \\ 
    \textbf{Model} & Acc@1 & Acc@5 & Acc@10 & MRR & Acc@1 & Acc@5 & Acc@10 & MRR \\ 
    \hline
    MF & 0.0352 & 0.0954 & 0.1499 & 0.0665 & 0.0239 & 0.0652 & 0.1289 & 0.0503\\ 
    LSTM & 0.1037 & 0.2076 & 0.2531 & 0.1563 & 0.1045 & 0.2167 & 0.2682 & 0.1616\\ 
    GSTN & 0.149 & 0.2913 & 0.3602 & 0.2171 & 0.1432 & 0.3045 & 0.3562 & 0.2234\\ 
    STGCN & 0.1843 & 0.3839 & 0.4472 & 0.2741 & 0.1579 & 0.3313 & 0.3965 & 0.2390\\ 
    PLSPL & 0.1959 & 0.4513 & 0.5394 & 0.2939 & 0.1812 & 0.3885 & 0.4763 & 0.2745\\ 
    STAN & 0.1842 & 0.4680 & 0.5846 & 0.3101 & 0.1720 & 0.3420 & 0.4201 & 0.2423\\ 
    CLSPR & 0.1884 & 0.4376 & 0.5529 & 0.2981 & 0.1773 & 0.4292 & 0.5316 & 0.3061\\ 
    GETNext & \underline{0.2482} & \underline{0.5031} & \underline{0.5905} & \underline{0.3618} & \underline{0.2254} & \underline{0.4417} & \underline{0.5287} & \underline{0.3267}\\ 
    \hline
    \textbf{GDPW} & \textbf{0.2634} & \textbf{0.5172} & \textbf{0.5910} & \textbf{0.3747} & \textbf{0.2511} & \textbf{0.4671} & \textbf{0.5304} & \textbf{0.3468}\\ 
    \toprule
\end{tabular}
\label{tab:results}
\end{table}

\subsection{Ablation Study}

We conduct an ablation study to demonstrate the impact of each component in our model on the overall prediction performance for the NYC and TKY datasets. The study consists of eight experimental configurations, as follows:
\begin{itemize}
    \item \textbf{w/o Category Graph}: We remove the Global Category Graph and replace it with category embeddings.
    \item \textbf{w/o Category-Time Graph}: We remove the Global Category-Time Graph and replace it with time embeddings.
    \item \textbf{w/o UG Graph}: We remove the Global UG Graph and replace it with POI embeddings.
    \item \textbf{w/o Contrastive}: The model that removed contrastive learning and \(L_{\text{CL}}\).
    \item \textbf{w/o Disentangle layer}: Without the whole Category-Time Disentangle Layer.
    \item \textbf{w/o TM}: Without the Transition Weighted Map.
    \item \textbf{w/o DM}: Without the Distance Map.
    \item \textbf{Change UG Graph}: Change the definition of Global UG graph edge weight from Newton's universal gravitation to the reciprocal distance.
    \item \textbf{w/o Category Prediction}: Without the Category Prediction part and \(L_{\text{cat}}\).
    \item \textbf{w/o Time Prediction}: Without the Time Prediction part and \(L_{\text{time}}\).
    \item \textbf{The full GDPW}.
\end{itemize}

The results are shown in Table \ref{tab:ablation}. The results indicate that removing the three global graphs respectively leads to a performance drop, with the most significant decrease observed when the Global UG Graph is omitted. This is because the Global UG Graph provides crucial information about the importance of each POI and learns the distance relationships between them. In contrast, removing the Global Category Graph and the Global Category-Time Graph had a minor impact, as these two graphs primarily provide auxiliary information. Additionally, the absence of contrastive learning and \(L_{\text{cat}}\) has a relatively minor impact on the model. The w/o TM and w/o DM scenarios demonstrate that weighting the POI transition weight and the distances between POIs can help improve performance. We also modified the edge weight definition in the UG graph and observed a slight performance drop. This indicates that using Newton’s law of universal gravitation to weight POI activities helps the model learn better. The w/o Category Prediction and w/o Time Prediction scenarios further demonstrate that Auxiliary Prediction benefits the model.

\begin{table}[h]
\caption{Results of ablation study.}
\centering
\begin{tabular}{ccccccccc}
\toprule
    \textbf{Dataset} & \multicolumn{4}{c}{\textbf{NYC}} & \multicolumn{4}{c}{\textbf{TKY}} \\ 
    \textbf{Model} & Acc@1 & Acc@5 & Acc@10 & MRR & Acc@1 & Acc@5 & Acc@10 & MRR \\ 
    \hline
    
    w/o Category graph & 0.2488 & 0.5160 & 0.5835 & 0.3646 & 0.2477 & 0.4567 & 0.5259 & 0.3424\\
    w/o Cat.-Time graph & 0.2591 & 0.5141 & 0.5875 & 0.3714 & 0.2503 & 0.4667 & 0.5287 & 0.3459\\
    w/o UG graph & 0.2423 & 0.4936 & 0.5679 & 0.3518 & 0.2285 & 0.4228 & 0.4875 & 0.3173\\
    w/o Contrastive & 0.2502 & 0.5145 & 0.5829 & 0.3653 & 0.2491 & 0.4584 & 0.5282 & 0.3443\\
    w/o Disentangle Layer & 0.2442 & 0.4801 & 0.5530 & 0.3499 & 0.2314 & 0.4296 & 0.4994 & 0.3233\\
    w/o TM & 0.2407 & 0.5025 & 0.5759 & 0.3546 & 0.2328 & 0.4447 & 0.5132 & 0.3284\\
    w/o DM & 0.2349 & 0.5040 & 0.5805 & 0.3523 & 0.2315 & 0.4436 & 0.5117 & 0.3265\\
    Change UG graph & 0.2545 & 0.5143 & 0.5889 & 0.3678 & 0.2503 & 0.4631 & 0.5302 & 0.3466\\
    w/o Category Prediction & 0.2468 & 0.5062 & 0.5813 & 0.3603 & 0.2482 & 0.4609 & 0.5300 & 0.3438\\
    w/o Time Prediction & 0.2501 & 0.5076 & 0.5834 & 0.3623 & 0.2478 & 0.4537 & 0.5210 & 0.3413\\
    \hline
    \textbf{GDPW} & \textbf{0.2634} & \textbf{0.5172} & \textbf{0.5910} & \textbf{0.3747} & \textbf{0.2511} & \textbf{0.4671} & \textbf{0.5304} & \textbf{0.3468}\\ 
    \toprule
\end{tabular}
\label{tab:ablation}
\end{table}

\subsection{Map Visualization}

\begin{figure}[h]
  \centering
  \includegraphics[width=\linewidth]{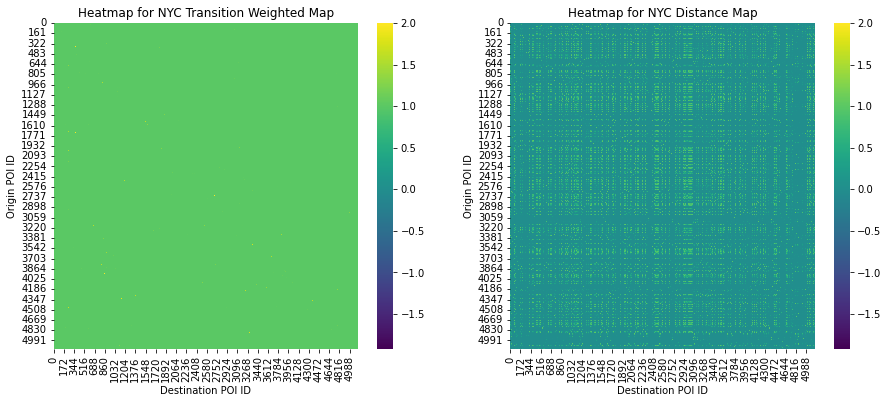}
  \caption{NYC Transition Weighted Map and Distance Map visualization.}
  \Description{NYC Transition Weighted Map and Distance Map visualization.}
  \label{fig:new_NYC}
\end{figure}

\begin{figure}[h]
  \centering
  \includegraphics[width=\linewidth]{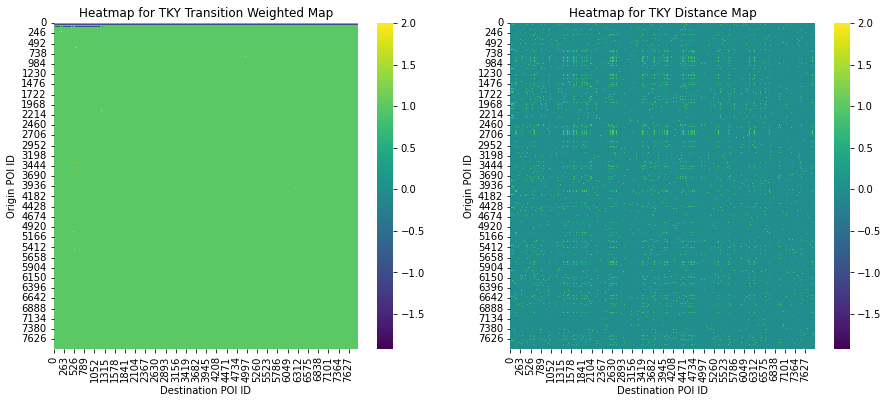}
  \caption{NYC Transition Weighted Map and Distance Map visualization.}
  \Description{NYC Transition Weighted Map and Distance Map visualization.}
  \label{fig:new_TKY}
\end{figure}

The Transition Weighted Map and the Distance Map each apply different weightings: one focuses on the transition weights between POIs, and the other on the distances between them. To illustrate how these maps weight different types of POIs, we use heatmaps to visualize the values in each map. As shown in Figure \ref{fig:new_NYC} and Figure \ref{fig:new_TKY}, the y-axis represents the starting POI ID, and the x-axis represents the destination POI ID. Each cell in the heatmap represents the transition weight or distance weight between any two POIs. We can observe that the Transition Weighted Map generally has a darker overall color, with only a few points being significantly more prominent. This is because the Transition Weighted Map applies weights to individual POIs. However, not all POIs have strong relationships, and only certain cells are weighted. In contrast, the Distance Map calculates the distances between POIs, resulting in a more evenly distributed color.

\begin{figure}[h]
  \centering
  \includegraphics[width=\linewidth]{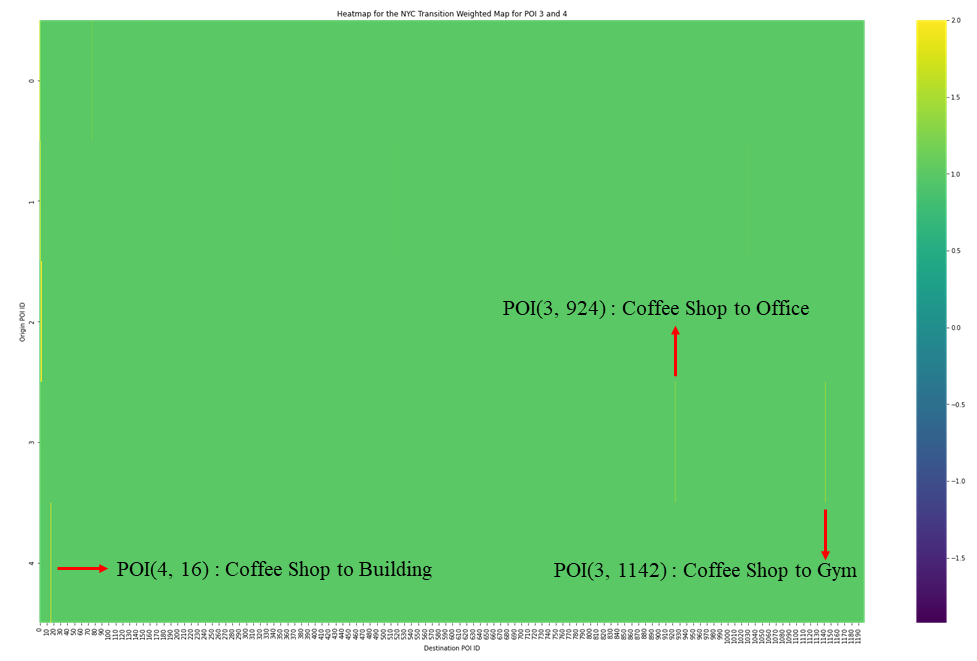}
  \caption{Heatmap for NYC Transition Weighted Map for POI 3 and POI 4.}
  \Description{Heatmap for NYC Transition Weighted Map for POI 3 and POI 4.}
  \label{fig:Heatmap for NYC Transition Weighted Map for POI 3 and POI 4}
\end{figure}

\begin{figure}[h]
  \centering
  \includegraphics[width=\linewidth]{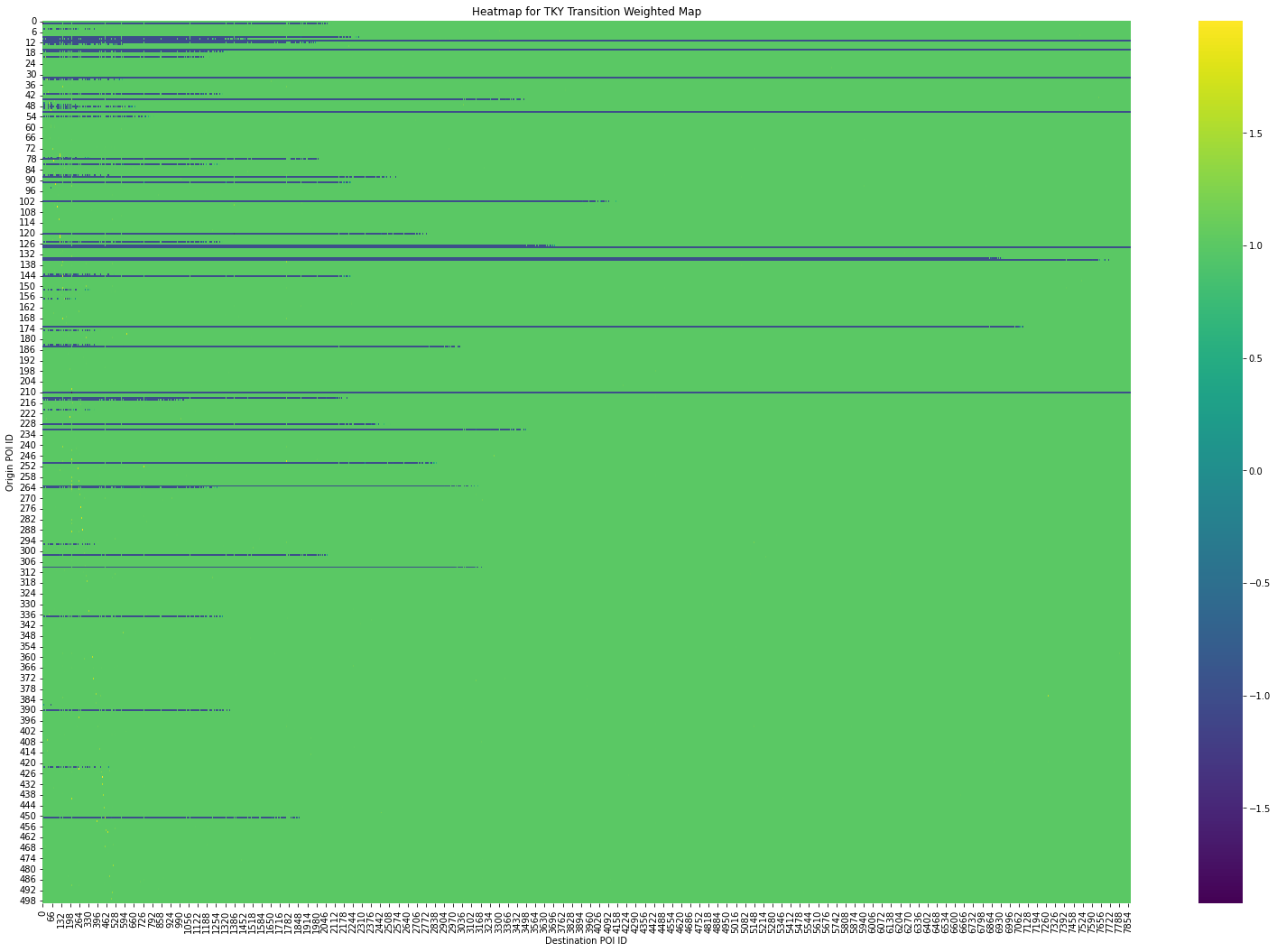}
  \caption{Heatmap for TKY Transition Weighted Map.}
  \Description{Heatmap for TKY Transition Weighted Map.}
  \label{fig:black_line}
\end{figure}

Due to the excessive number of cells in the figure, it is difficult to easily observe the POIs weighted in the Transition Map (TM). Therefore, we enlarge the TM heatmap for analysis. The enlarged images of the NYC TM and TKY TM are shown in Figure \ref{fig:Heatmap for NYC Transition Weighted Map for POI 3 and POI 4} and Figure \ref{fig:black_line}, respectively. In NYC, we take POI 3 (Food Truck) and POI 4 (Coffee Shop) as examples. For POI 3, the points with a transition weight greater than 1 are POI 924 (Office) and POI 1142 (Gym). For POI 4, the prominent transition behavior is to POI 16 (Building), indicating that the NYC TM mainly captures everyday activities. However, in the Tokyo TM, we find that certain POIs have smaller weight values than other POIs. These points correspond to POI 11 (Ikebukuro Train Station), POI 16 (Tokyo Train Station), POI 32 (Shinjuku Train Station), POI 51 (Shibuya Train Station), POI 128 (Akihabara Train Station), and POI 210 (Shinagawa Train Station). Taking POI 16 as an example, we find that the POI pairs with a transition weight greater than 1 are (16, 11), (16, 32), and (16, 128). These pairs indicate that weighting is applied between major stations. This suggests that check-ins between major stations far exceed check-ins between major stations and other POIs, resulting in a lower relative weight for major stations to other POIs.

\subsection{Hyper-Parameter Study}

\begin{figure}[h]
  \centering
  \includegraphics[width=\linewidth]{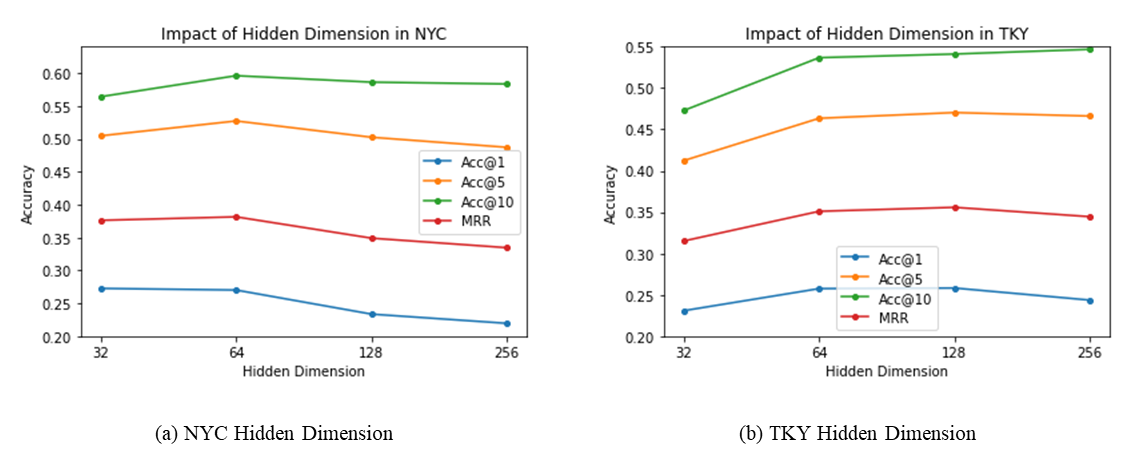}
  \caption{Impact of hidden dimensions.}
  \Description{Impact of hidden dimensions.}
  \label{fig:Hidden_dim}
\end{figure}

\begin{figure}[h]
  \centering
  \includegraphics[width=\linewidth]{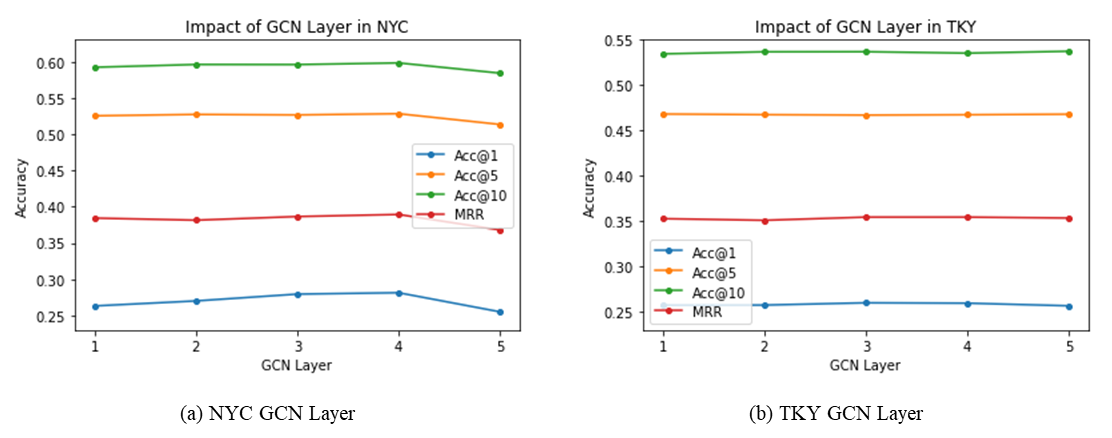}
  \caption{Impact of GCN layers.}
  \Description{Impact of GCN layers.}
  \label{fig:GCN_Layer}
\end{figure}

To ensure the stability of our model, we perform hyper-parameter experiments. We set the dimensions of the hidden layers \(d\) in the model to \([32, 64, 128, 256]\) and test the impact of different hidden layer dimensions on the model's performance. As shown in Figure \ref{fig:Hidden_dim}, in NYC, the best performance is achieved when the hidden dimension is set to 64. In TKY, the performance is comparable when the hidden dimension is 64 or 128. Considering runtime efficiency, we set the hidden dimension \(d = 64\). We also test the number of attention layers \(l\) in the model, as shown in Figure \ref{fig:GCN_Layer}. We found that the number of GCN layers has little impact on performance. Considering computational efficiency, we set the number of GCN layers to 2.

\section{Conclusions and Future Work}
\label{sec:conclusion}
\subsection{Conclusions}

This paper proposes a novel next POI recommendation model, GDPW, which is based on three global graphs. Specifically, we construct the Global Category Graph and the Global Category-Time graph, utilizing contrastive learning to disentangle category and time information. This approach helps the model to more comprehensively learn the relationships between different categories and times. Additionally, we propose a POI Weighted Layer to weight the number of check-ins of the POI itself, the transition weight, and the distance between POIs. Finally, the learned information is combined, and after prediction, the Transition Weighted Map and the Distance Map, established in the POI Weighted Layer, are used to weight the results. Through this method, the model can learn different aspects of information and make more accurate next POI recommendations. Experiments on a real-world dataset demonstrate that our model significantly outperforms all competitors.
\subsection{Future Work}

In future work, we plan to use different GNN models for each graph to better learn the representations within each graph. Additionally, GDPW currently uses concatenation or summation for fusion, but the fusion method can significantly impact performance. We believe that exploring different methods for fusing these various representations will be another direction worth investigating.
\begin{acks}
This work was partially supported by National Science and Technology Council (NSTC) under Grants 111-2636-E-006 -026 -, 112-2221-E-006 -100 - and 112-2221-E-006 -150 -MY3. 
The authors are grateful to Tainan City Government for providing the ovitrap data.
\end{acks}

\bibliographystyle{ACM-Reference-Format}
\bibliography{GDPW}
\end{document}